\documentclass[sigconf]{acmart}

\usepackage{xspace}
\usepackage[scaled=0.9]{helvet}
\usepackage{graphicx}
\usepackage{bmpsize}
\usepackage{paralist}
\usepackage{xcolor}
\usepackage{subcaption}
\usepackage{balance}
\usepackage{booktabs}
\usepackage{tabularx}
\usepackage{diagbox}
\usepackage[utf8]{inputenc}
\usepackage{makecell}
\usepackage{hyperref}
\usepackage{titlesec}
\usepackage{enumitem}
\usepackage{listings}
\usepackage{xcolor}
\usepackage{cleveref}

\newcommand{\lst}{\textsc{LST}\xspace}
\newcommand{\lsts}{\textsc{LSTs}\xspace}
\newcommand{\onetable}{\textsc{XTable}\xspace}
\newcommand{\smallsection}[1]{\vspace{1mm}\noindent\textbf{#1.}}

\definecolor{codegreen}{rgb}{0,0.6,0}
\definecolor{codegray}{rgb}{0.5,0.5,0.5}
\definecolor{codepurple}{rgb}{0.58,0,0.82}
\definecolor{backcolour}{rgb}{0.95,0.95,0.92}

\lstdefinestyle{mycodestyle}{
    commentstyle=\color{codegreen},
    keywordstyle=\color{magenta},
    stringstyle=\color{codepurple},
    basicstyle=\ttfamily\footnotesize,
    breakatwhitespace=false,         
    breaklines=true,                 
    captionpos=b,                    
    keepspaces=true,                 
    numbers=left,                    
    numberstyle=\color{codegray},
    xleftmargin=2em,
    frame=lines,
    framexleftmargin=2em,
    numbersep=5pt,                  
    showspaces=false,                
    showstringspaces=false,
    showtabs=false,                  
    tabsize=2
}

\title{\onetable in Action: Seamless Interoperability in Data Lakes}

\author{Ashvin Agrawal$^{1}$, Tim Brown$^{2}$, Anoop Johnson$^3$,\\ Jes\'us Camacho-Rodr\'iguez$^1$, Kyle Weller$^2$, Carlo Curino$^1$, Raghu Ramakrishnan$^1$}

\affiliation{
    \institution{$^1$Microsoft, $^2$Onehouse, $^3$Google}
    \country{}
}
\affiliation{
    $^1$\{ashvin.agrawal, jesusca, carlo.curino, raghu\}@microsoft.com, $^2$\{tim, kyle\}@onehouse.ai, $^3$anoopkj@google.com
    \country{}
}

\date{November 2023}

\begin{document}

\begin{abstract}

Contemporary approaches to data management are increasingly relying on unified analytics and AI platforms to foster collaboration, interoperability, seamless access to reliable data, and high performance.
\emph{Data Lakes} featuring open standard table formats such as Delta Lake, Apache Hudi, and Apache Iceberg are central components of these data architectures. 
Choosing the right format for managing a table is crucial for achieving the objectives mentioned above.
The challenge lies in selecting the best format, a task that is onerous and can yield temporary results, as the ideal choice may shift over time with data growth, evolving workloads, and the competitive development of table formats and processing engines.
Moreover, restricting data access to a single format can hinder data sharing resulting in diminished business value over the long term.
The ability to seamlessly interoperate between formats and with negligible overhead can effectively address these challenges.
Our solution in this direction is an innovative omni-directional translator, \onetable,
that facilitates writing data in one format and reading it in any format,
thus achieving the desired format interoperability. In this work, we
demonstrate the effectiveness of \onetable through application
scenarios inspired by real-world use cases.

\end{abstract}

\pagestyle{plain} 

\maketitle

\section{Introduction}
\label{sec:intro}

In the quest to harness the full potential of data, organizations are invariably dedicated to eliminating data silos and enabling analytics on a single copy of data, regardless of where or how the data is stored. 
Their goal is to not only guarantee uniform access to data for both proprietary and open-source analytical engines, but also to achieve seamless interoperability, enabling any engine to utilize the information processed in any format.
Simultaneously, organizations prioritize key aspects such as security, performance, management, and governance, all of which are significantly simplified by maintaining a single copy of data.  
Notable examples of platforms facilitating this integration include Google BigLake~\cite{biglake}, Microsoft Fabric~\cite{msfabric}, and Databricks~\cite{databricks}. 

At the core of this strategy lies a \emph{unified data lake}, which is the central data repository in these platforms. 
The data lake is designed to handle large volumes and diverse types of enterprise data, providing a scalable and secure environment. 
It allows data to be ingested from various sources, including cloud, on-premises, and edge-computing platforms. 
Additionally, it facilitates analytical and decision-making processes by ensuring data is easily accessible in formats usable by a wide range of applications like data engineering, data science, real-time analytics, and business intelligence. 

Conventional text formats like CSV and JSON, though widely used for data storage, prove inadequate when dealing with the vast scale of big data, rendering them suboptimal for data lakes. 
As a response to the escalating volumes of data, file formats such as Apache Avro~\cite{avro}, Apache Parquet~\cite{parquet}, and Apache ORC~\cite{orc} have emerged. 
These formats have gained prominence due to their optimizations for data storage and retrieval, especially in scenarios characterized as read-heavy. 
It is worth noting that these formats are designed to be immutable. 
However, modern analytics scenarios do not just demand static analysis of large data sets; they also require frequent, incremental updates to structured data in small batches---a demand these formats are not suited to meet. 

In this context, Log-Structured Tables (\lsts) have emerged as compelling proposals to address the aforementioned challenges. 

\lsts are table storage formats that combine the advantages of optimized file formats with a design suited for frequent table updates (\S\ref{sec:lstoverview}). They have evolved into an industry standard; 
notable implementations that have gained widespread adoption include Delta Lake~\cite{delta-lake}, Apache Iceberg~\cite{apache-iceberg}, and Apache Hudi~\cite{apache-hudi}. 
Each \lst adopts a distinct approach to storing and managing metadata, including versioning, schema, and partitioning. 
This diversity arises from the fact that each \lst is developed by open-source communities with different goals, use cases, and requirements, leading to unique features and capabilities. 

To achieve interoperability and unified experiences across engines, organizations must decide between standardizing on a single \lst for their entire data lake~\cite{iceberg-at-apple,hudi-at-walmart,delta-lake-at-fabric} or opting for engines that connect to many \lsts~\cite{starburst-connectors}. 
However, both strategies involve a trade-off; the performance varies significantly across different workload types, and tends to fluctuate over time with changes in the characteristics of the workloads, \lsts, infrastructure, and the engines~\cite{lstbench, lhbench}. 
This suggests that relying exclusively on a single \lst or engine may not be fruitful always. 
The problem of data estate fragmentation is only partially addressed without solutions that provide interoperability in both engine and format.
Consequently, users often resort to data duplication, posing a challenge to the overarching goal of achieving unification and interoperability. 
This duplication not only raises costs significantly but can also decrease the \emph{value} of the data, e.g., the timeliness of insights and actions based on the ingested data. 

To leverage the advantages offered by different \lsts while adhering to the data lake vision, \emph{\lst translation} has emerged as an effective approach. 
The translation involves generating metadata for one \lst from that of another, while reusing the data files.
This means that the data, written in a format, is simultaneously available in other formats.
Given that the table metadata is considerably smaller in size and the translation does not involve larger data files, this process is both low-overhead and swift.
\lst translation preserves the `single source of truth' concept while simultaneously enhancing format interoperability.

In this direction, we introduce \onetable~\cite{oss-onetable} (\S\ref{sec:onetable}), a tool designed to realize seamless interoperability in data lakes. 
\onetable is an \emph{open-sourced} tool providing omni-directional and incremental translation between \lsts, operates with low overheads and high performance, and is designed for extensibility and usability. 
In this demonstration, we employ \onetable to target diverse application scenarios and user requirements, showcasing its versatility and capabilities.
The participants will (1)~use notebooks to explore the structure of data and metadata layout of various table formats, examining their differences and similarities, (2)~engage with unified analytical platforms to delve into query planning using table format features, and (3)~have the opportunity to interact with \onetable and experience its practical applications firsthand through exercises inspired by real-world scenarios.

\section{\lst Overview}
\label{sec:lstoverview}

A Log-Structured Table (\lst or simply a table) is a specification designed for storing versioned tabular data 
using optimized file formats such as Parquet and ORC. The specifications guarantee data integrity and consistency 
through adherence to ACID transactions. An \lst is primarily composed of data files and metadata files.

The data files store the table's records and are typically organized according to the table's partitioning scheme. 
Once written, data files are immutable, i.e., any change to the \lst data results in the creation of new data files. 
On the other hand, the metadata files contain details about the table, including its schema, partition scheme, location of 
data files, file-level statistics, and configuration. \lsts use the metadata files for version control, where 
each commit operation is reflected through a subset of metadata files that capture the table's state. 
The metadata files associated with a new commit contain references to both newly added and existing data files, 
while ensuring that no existing files are deleted. This ensures completion of concurrent read operations successfully.
The metadata update process is designed to be atomic. This design also supports features like schema evolution and 
time travel (querying historical data). The separation of data and metadata facilitates efficient 
management of big datasets, as task planning deals with smaller and fewer metadata files, and is independent of the numerous larger data files.

\begin{lstlisting}[float, language=sql, caption=Example code for LST operations, label=lst:iceberg_sample]
CREATE TABLE sales (s_id int, s_type string) USING ICEBERG PARTITIONED BY (s_type);
INSERT INTO sales VALUES (1, 'a'), (2, 'b'), (3, 'b');
DELETE FROM sales WHERE s_id = 3;
\end{lstlisting}

\begin{figure}[t]
    \centering
    \includegraphics[scale=0.70]{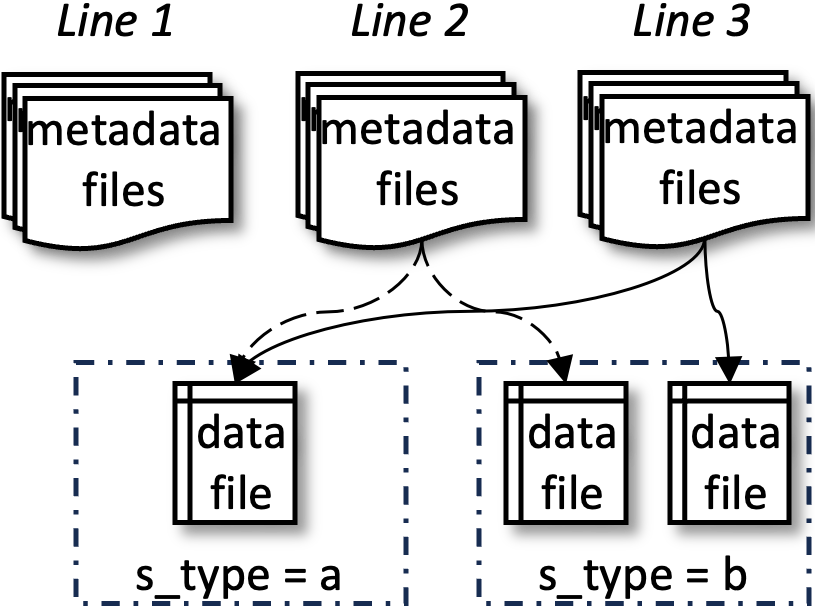}\vspace{-0.5em}
    \caption{Simplified layout of metadata files and referenced data files for example in ~\Cref{lst:iceberg_sample}}\vspace{-0.5em}
    \label{fig:iceberg_eg_layout}
\end{figure}

\Cref{lst:iceberg_sample} and \Cref{fig:iceberg_eg_layout} serves as an example to demonstrate working of a \lst named \emph{sales} using Apache Iceberg. 
On \emph{Line 1}, the table is created with two columns: \emph{s\_id} and \emph{s\_type}. As it is partitioned, Iceberg 
will arrange table's data files in folders according to values of \emph{s\_type}. 
At this stage, initial metadata files are generated, capturing the table's schema, partition details, and no data files.
On \emph{Line 2}, three rows are added to \emph{sales}. Iceberg processes this by creating one or more data files. 
Following this, metadata files are created, which only reference the newly added data files. 
\emph{Line 3} removes one record. Iceberg generates new data files from existing files (assuming copy-on-write mode),
omitting the deleted records. Subsequently, new metadata files reflecting the post-deletion table state are created. 
This latest metadata files do not refer to any data files that contained the now-deleted records. 
Together, the three sets of metadata files represent the table's history.

Interestingly, Delta Lake, Iceberg, and Hudi are designed to be compatible with standard file formats, and their data file layout typically aligns with the previously described layout. The primary distinction lies in their metadata layers, where the differences are minor. The metadata across these three formats fulfills similar functions and contains similar types of information.

Despite having similar metadata information, real-world performance of \lsts can vary dramatically~\cite{lstbench, lhbench}.
The metadata layout is just one of many factors that contribute to performance variations. Other responsible factors include the characteristics of the workload, the specific algorithms implemented within the \lst, and the protocols governing engine interactions which affect concurrency and isolation. 
Furthermore, configuration parameters play a significant role, influencing broader aspects such as the data file sizes. 
Consequently, to maximize the performance by fully utilizing the potential of \lsts, users might find it necessary to use different engines for different tasks, with each engine supporting multiple formats. This approach then allows them to select the best fit based on their workload and environment.

A hurdle in attaining such interoperability lies in the continuous evolution of the formats and the processing engines. This necessitates a continual update of the connectors, which is difficult to maintain. This leads to varying levels of support in these integrations, such as the lack of uniform optimization support, resulting in situations where optimizations effective for one format might not be applicable to another~\cite{trino-hudi-perf}.

\section{\onetable Overview}
\label{sec:onetable}

One of the most efficient strategies for enabling interoperability is to make the data available in the preferred format, rather than maintaining and extending all engines.
\onetable is an innovative \emph{omni-directional}, \emph{incremental}, \emph{extensible}, and \emph{low-overhead} translator providing 
seamless interop between \lsts. It facilitates scenarios where data, created in a \emph{source \lst}, becomes accessible across 
all supported \emph{target \lsts} without data duplication. Significantly, \onetable is not a new \lst itself and does not interact with engines directly. 
It operates asynchronously, ensuring that existing applications optimized for specific \lst can seamlessly work with both original and translated data, eliminating the need for maintaining multiple connectors.
The core underlying principle of \onetable is the realization that the differences in the metadata layers of the currently popular \lsts are relatively minor. 

\smallsection{Omni-directional} \onetable has the ability to translate metadata in any direction. It means that it doesn’t matter what the source or target \lst is; \onetable can handle the translation both ways. 
With its unique omni-directional conversion capability, infrastructure teams are relieved from the burden of deploying and maintaining separate translators and connectors for each table format.

\smallsection{Low-overhead} \lst translation is primarily focused on the metadata, which is a significantly lighter task compared to processing large data files. In most cases, the actual data files do not need to be read for the translation to occur. Furthermore, since the differences in the metadata between various \lsts are minor, the translation process does not demand extensive computational resources.

\smallsection{Incremental} Incremental translation enhances \onetable's efficiency. \onetable can 
detect which source \lst commits have not yet been translated to the target \lst and focuses solely 
on converting those. This approach significantly reduces the time and required computational 
resources. This allows for frequent invocations of \onetable, even on a commit-by-commit basis, effectively 
minimizing the staleness of the target data.

\smallsection{Extensible} \onetable is designed to easily adapt and expand in response to emerging table formats~\cite{apache-paimon} 
or new versions of existing \lsts. This adaptability is enabled by an internal representation that serves as a universal
exchange mechanism, effectively bridging different formats. The internal representation plays a pivotal role in 
simplifying development and validation by effectively isolating source formats from target formats. This isolation 
means that when adding support for a new format or updating an existing one, developers can focus solely on how to
translate to and from this internal representation, rather than dealing with the complexities of direct format-to-format translation. This design fosters community-led development of \onetable and quickly assimilate new formats and updates.

\begin{lstlisting}[float, caption=Example \onetable config, label=lst:onetable_sample]
sourceFormat: HUDI
targetFormats:
  - DELTA
  - ICEBERG
datasets:
  -
    tableBasePath: abfs://container@ac.dfs.core.windows.net/sales
\end{lstlisting}

\Cref{lst:onetable_sample} is an example \onetable configuration file. 
\onetable requires three key inputs: the source \lst (Hudi in this case), a list of target \lsts (Delta and Iceberg), and the location of the data in the source \lst. 
The specified location is an Azure Blob File System (ABFS) path, which hosts the \emph{sales} table in Hudi format. \onetable generates Delta-Lake and Iceberg metadata from the Hudi metadata. The generated metadata is persisted in the base path of \emph{sales} table colocated with the data files.

\subsection{\onetable Architecture}
\begin{figure}[t]
    \centering
    \includegraphics[scale=0.70]{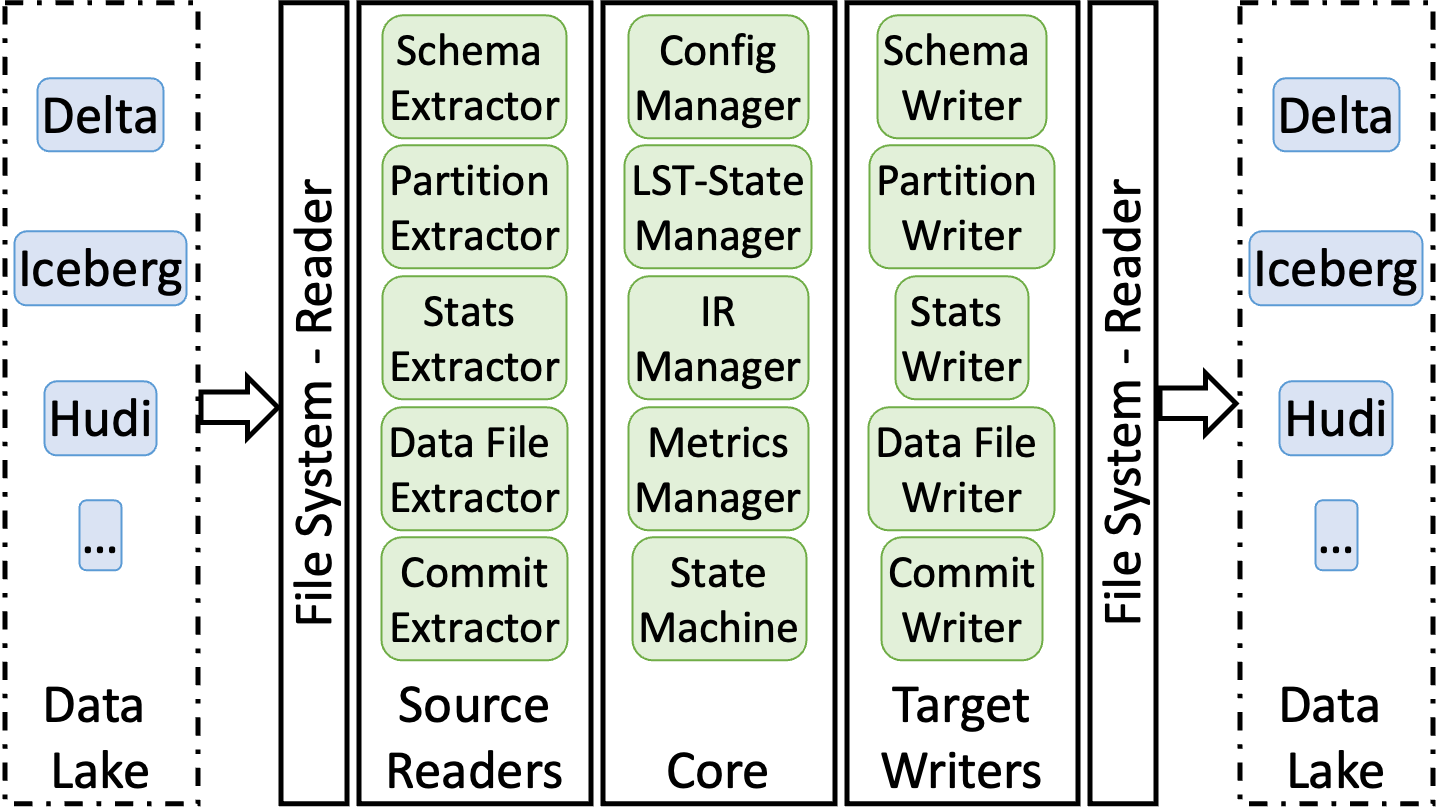}\vspace{-0.5em}
    \caption{Overview of the \onetable Architecture.}\vspace{-0.5em}
    \label{fig:onetable_arch}
\end{figure}
At a high level, as illustrated in the ~\Cref{fig:onetable_arch}, \onetable's architecture is comprised of three key components, the source readers, the target writers, and the central core logic.

\smallsection{Source Readers} These are \lst specific modules responsible for reading metadata from the source tables. 
They operate using a pluggable file system, allowing them to connect to different data lake implementations. 
The source readers extract information like schema, transactions, and partitions, and translate it into \onetable's unified internal representation. 

\smallsection{Target Writers} These mirror the source readers. Their role is to take the internal representation of the metadata
and accurately map it to the target format’s metadata structure. This includes recreating schema, transaction logs, and partition details in the new format.

\smallsection{Core Logic} This is the central processing unit of \onetable. 
It orchestrates the entire translation process, including initializing of all components, managing sources and targets, 
and handling tasks like caching for efficiency, state management for recovery and incremental processing, 
and telemetry for monitoring.

\section{Related Work}
\label{sec:related}

Delta's \emph{UniForm}~\cite{delta-uniform} and 
Iceberg's \emph{Migrate Table}~\cite{iceberg-migrate} are related technologies that, while addressing interoperability challenges, offer translation in only one direction compared to \onetable. UniForm specifically focuses on efficient, 
incremental translation from Delta metadata to Iceberg. It incurs minimal impact on Delta write 
performance, as the transaction is performed after the Delta commit.
Similarly, the \emph{Migrate} action creates a new Iceberg table based on a specific point-in-time snapshot
of a source table. However, this process also removes the source table from the catalog, eliminating
the option to using the original format.
While the two processes share core principles with \onetable, further validating the approach's
effectiveness, their capabilities are more focused. This makes it a useful yet narrower solution in the  context of \lst translation.

\section{Demonstration Overview}
\label{sec:demo}

The demonstration, grounded in realistic use cases, is designed to help the attendees understand how \onetable's omni-directional \lst translation capabilities enhance interoperability and efficiency in data lakes.
To this end, we deploy \onetable as a background process which is triggered asynchronously either
periodically or on demand following one or more commit operations by analytical engines such as Apache Spark, Trino, and Apache Flink.
The engines and \onetable do not interact, eliminating need for code changes or special engine
configurations.

\smallsection{Utilities Package} \onetable demonstration comprises of two main packages. The first is a set of utilities designed to
visualize and examine artifacts generated by \onetable, \lsts, and query engines. The utilities
include python notebooks (1) to visualize file layout and the structure of key metadata files of the
\lsts to highlight the working of the formats and compare them with each other, (2) examine
execution plans of queries submitted by attendees, and (3) visualize the timeline view of \onetable
events and the work done.
Attendees will interact with these python notebooks throughout the exercise scenarios discussed below.

\smallsection{Applications Package} The second package consists of practical exercises, enabling attendees to engage directly with real-world applications of \onetable and experience its impact firsthand.
These exercises have been designed using publicly available datasets and a mix of open-sourced and proprietary analytical engines. Our demonstration assigns attendees different roles, such as a data engineer, performance engineer, and a data scientist, enabling them to unlock engine and format interoperability on a single copy of data using \onetable.

\smallsection{Scenario 1: Multi-Format Data Import/Export in Managed Environments}
The prevailing approach for analytics and AI in managed environments is to stick to a single \lst. 
Yet, in practice, when collaborating with external partners who use different formats, numerous organizations
depend on various data sources and often distribute their data in multiple formats. 
In this scenario, an e-commerce company uses a managed data lake for its varied data storage and analysis needs.
In the role of a data engineer, the attendee will complete data integration exercises by specifying the locations of datasets and using \onetable to import multi-format data from partners and export their data in the format required by their partners.
This capability practically unlocks their data and eliminates the need to employ ad-hoc processes to managing multiple data copies in different formats.

\begin{figure}[t]
    \centering
    \includegraphics[scale=0.55]{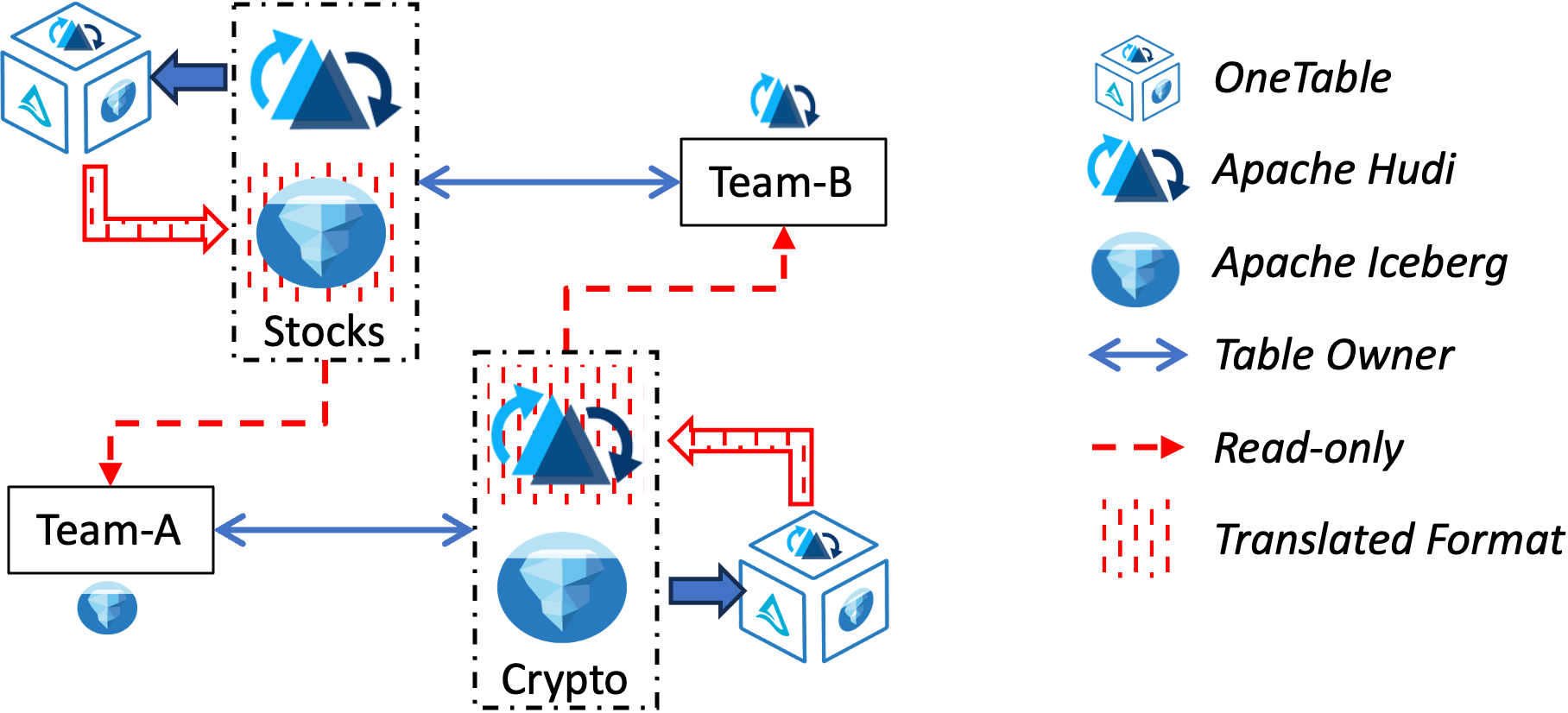}\vspace{-0.5em}
    \caption{Scenario 2: \emph{Team-A} uses Iceberg exclusively, and \emph{Team B} uses Hudi. \onetable enables access to \emph{Stocks} and \emph{Crypto} tables in both formats for seamless interop.}\vspace{-0.5em}
    \label{fig:demo-s2}
\end{figure}

\smallsection{Scenario 2: Unlocking Analytics Across Table Formats}
In a global financial firm, different teams use different analytics engines and \lsts for their operations. 
Team A processes transactional data using Apache Iceberg, while Team B prefers Apache Hudi for their market analysis. 
Typically, sharing insights across these teams requires extensive data conversion and coordination. 
With \onetable, both teams can access and analyze most up-to-date version of each other's data seamlessly 
without the need for time-consuming coordination efforts.
As a data scientist of Team A, the attendee will use their preferred stack, Spark and Iceberg, to analyze Team B's Hudi-formatted \emph{Stock} table's data directly for completing a forecasting exercise, as shown in ~\Cref{fig:demo-s2}.

\smallsection{Scenario 3: Optimizing Performance with Engine Flexibility}
In this scenario, a healthcare organization collects vast amounts of sensor data in streaming fashion in 
Hudi-based tables. For certain complex queries, they find their analytical engine, Trino, is optimized for using column 
statistics in Iceberg, offering faster query execution and efficient data processing. 
Using \onetable, the attendee, acting as a performance engineer, will convert their Hudi-based datasets to Iceberg, utilize Trino for certain complex queries. This flexibility allows them to choose the best engine for the task at hand without duplicating data or
compromising on data integrity. The attendee will leverage the query plan visualization tool from the utilities package for this exercise.

\bibliographystyle{ACM-Reference-Format}
\bibliography{ref}


\begin{thebibliography}{20}


\ifx \showCODEN    \undefined \def \showCODEN     #1{\unskip}     \fi
\ifx \showDOI      \undefined \def \showDOI       #1{#1}\fi
\ifx \showISBNx    \undefined \def \showISBNx     #1{\unskip}     \fi
\ifx \showISBNxiii \undefined \def \showISBNxiii  #1{\unskip}     \fi
\ifx \showISSN     \undefined \def \showISSN      #1{\unskip}     \fi
\ifx \showLCCN     \undefined \def \showLCCN      #1{\unskip}     \fi
\ifx \shownote     \undefined \def \shownote      #1{#1}          \fi
\ifx \showarticletitle \undefined \def \showarticletitle #1{#1}   \fi
\ifx \showURL      \undefined \def \showURL       {\relax}        \fi
\providecommand\bibfield[2]{#2}
\providecommand\bibinfo[2]{#2}
\providecommand\natexlab[1]{#1}
\providecommand\showeprint[2][]{arXiv:#2}

\bibitem[\protect\citeauthoryear{??}{apa}{2017}]%
        {apache-hudi}
 \bibinfo{year}{2017}\natexlab{}.
\newblock \bibinfo{title}{Apache Hudi}.
\newblock \bibinfo{howpublished}{\url{https://hudi.apache.org/}}.
\newblock
\newblock
\shownote{Accessed: 2023-02-23.}


\bibitem[\protect\citeauthoryear{??}{del}{2019}]%
        {delta-lake}
 \bibinfo{year}{2019}\natexlab{}.
\newblock \bibinfo{title}{Delta Lake}.
\newblock \bibinfo{howpublished}{\url{https://delta.io/}}.
\newblock
\newblock
\shownote{Accessed: 2023-02-23.}


\bibitem[\protect\citeauthoryear{??}{apa}{2021}]%
        {apache-iceberg}
 \bibinfo{year}{2021}\natexlab{}.
\newblock \bibinfo{title}{Apache Iceberg}.
\newblock \bibinfo{howpublished}{\url{https://iceberg.apache.org/}}.
\newblock
\newblock
\shownote{Accessed: 2023-02-23.}


\bibitem[\protect\citeauthoryear{??}{avr}{2023}]%
        {avro}
 \bibinfo{year}{2023}\natexlab{}.
\newblock \bibinfo{title}{Apache Avro}.
\newblock \bibinfo{howpublished}{\url{https://avro.apache.org/}}.
\newblock
\newblock
\shownote{Accessed: 2023-02-23.}


\bibitem[\protect\citeauthoryear{??}{orc}{2023}]%
        {orc}
 \bibinfo{year}{2023}\natexlab{}.
\newblock \bibinfo{title}{Apache ORC}.
\newblock \bibinfo{howpublished}{\url{https://orc.apache.org/}}.
\newblock
\newblock
\shownote{Accessed: 2023-02-23.}


\bibitem[\protect\citeauthoryear{??}{apa}{2023}]%
        {apache-paimon}
 \bibinfo{year}{2023}\natexlab{}.
\newblock \bibinfo{title}{Apache Paimon}.
\newblock \bibinfo{howpublished}{\url{https://paimon.apache.org}}.
\newblock
\newblock
\shownote{Accessed: 2023-11-29.}


\bibitem[\protect\citeauthoryear{??}{par}{2023}]%
        {parquet}
 \bibinfo{year}{2023}\natexlab{}.
\newblock \bibinfo{title}{Apache Parquet}.
\newblock \bibinfo{howpublished}{\url{https://parquet.apache.org/}}.
\newblock
\newblock
\shownote{Accessed: 2023-02-23.}


\bibitem[\protect\citeauthoryear{??}{ice}{2023a}]%
        {iceberg-at-apple}
 \bibinfo{year}{2023}\natexlab{a}.
\newblock \bibinfo{title}{Apple-Iceberg}.
\newblock \bibinfo{howpublished}{\url{https://trino.io/blog/2022/11/28/trino-summit-2022-apple-recap.html}}.
\newblock
\newblock
\shownote{Accessed: 2023-11-29.}


\bibitem[\protect\citeauthoryear{??}{dat}{2023}]%
        {databricks}
 \bibinfo{year}{2023}\natexlab{}.
\newblock \bibinfo{title}{Databricks}.
\newblock \bibinfo{howpublished}{\url{https://www.databricks.com/}}.
\newblock
\newblock
\shownote{Accessed: 2023-11-26.}


\bibitem[\protect\citeauthoryear{??}{del}{2023a}]%
        {delta-lake-at-fabric}
 \bibinfo{year}{2023}\natexlab{a}.
\newblock \bibinfo{title}{Fabric Interoperability}.
\newblock \bibinfo{howpublished}{\url{https://learn.microsoft.com/en-us/fabric/get-started/delta-lake-interoperability}}.
\newblock
\newblock
\shownote{Accessed: 2023-11-29.}


\bibitem[\protect\citeauthoryear{??}{big}{2023}]%
        {biglake}
 \bibinfo{year}{2023}\natexlab{}.
\newblock \bibinfo{title}{Google BigLake}.
\newblock \bibinfo{howpublished}{\url{https://cloud.google.com/biglake}}.
\newblock
\newblock
\shownote{Accessed: 2023-11-26.}


\bibitem[\protect\citeauthoryear{??}{msf}{2023}]%
        {msfabric}
 \bibinfo{year}{2023}\natexlab{}.
\newblock \bibinfo{title}{Microsoft Fabric}.
\newblock \bibinfo{howpublished}{\url{https://learn.microsoft.com/en-us/fabric/}}.
\newblock


\bibitem[\protect\citeauthoryear{??}{ice}{2023b}]%
        {iceberg-migrate}
 \bibinfo{year}{2023}\natexlab{b}.
\newblock \bibinfo{title}{Migrate Action}.
\newblock \bibinfo{howpublished}{\url{https://iceberg.apache.org/docs/1.3.0/table-migration/}}.
\newblock
\newblock
\shownote{Accessed: 2023-12-07.}


\bibitem[\protect\citeauthoryear{??}{oss}{2023}]%
        {oss-onetable}
 \bibinfo{year}{2023}\natexlab{}.
\newblock \bibinfo{title}{OneTable}.
\newblock \bibinfo{howpublished}{\url{https://onetable.dev}}.
\newblock
\newblock
\shownote{Accessed: 2023-11-29.}


\bibitem[\protect\citeauthoryear{??}{sta}{2023}]%
        {starburst-connectors}
 \bibinfo{year}{2023}\natexlab{}.
\newblock \bibinfo{title}{Starburst}.
\newblock \bibinfo{howpublished}{\url{https://docs.starburst.io/latest/object-storage.html}}.
\newblock
\newblock
\shownote{Accessed: 2023-11-29.}


\bibitem[\protect\citeauthoryear{??}{tri}{2023}]%
        {trino-hudi-perf}
 \bibinfo{year}{2023}\natexlab{}.
\newblock \bibinfo{title}{Trino Hudi Connector Issue}.
\newblock \bibinfo{howpublished}{\url{https://github.com/trinodb/trino/pull/17899}}.
\newblock
\newblock
\shownote{Accessed: 2023-12-07.}


\bibitem[\protect\citeauthoryear{??}{del}{2023b}]%
        {delta-uniform}
 \bibinfo{year}{2023}\natexlab{b}.
\newblock \bibinfo{title}{Universal Format (UniForm)}.
\newblock \bibinfo{howpublished}{\url{https://learn.microsoft.com/en-us/azure/databricks/delta/uniform}}.
\newblock
\newblock
\shownote{Accessed: 2023-06-23.}


\bibitem[\protect\citeauthoryear{??}{hud}{2023}]%
        {hudi-at-walmart}
 \bibinfo{year}{2023}\natexlab{}.
\newblock \bibinfo{title}{Walmart-Hudi}.
\newblock \bibinfo{howpublished}{\url{https://medium.com/walmartglobaltech/lakehouse-at-fortune-1-scale-480bcb10391b/}}.
\newblock
\newblock
\shownote{Accessed: 2023-12-08.}


\bibitem[\protect\citeauthoryear{Camacho-Rodríguez, Agrawal, Gruenheid, Gosalia, Petculescu, Aguilar-Saborit, Floratou, Curino, and Ramakrishnan}{Camacho-Rodríguez et~al\mbox{.}}{2023}]%
        {lstbench}
\bibfield{author}{\bibinfo{person}{Jesús Camacho-Rodríguez}, \bibinfo{person}{Ashvin Agrawal}, \bibinfo{person}{Anja Gruenheid}, \bibinfo{person}{Ashit Gosalia}, \bibinfo{person}{Cristian Petculescu}, \bibinfo{person}{Josep Aguilar-Saborit}, \bibinfo{person}{Avrilia Floratou}, \bibinfo{person}{Carlo Curino}, {and} \bibinfo{person}{Raghu Ramakrishnan}.} \bibinfo{year}{2023}\natexlab{}.
\newblock \bibinfo{title}{LST-Bench: Benchmarking Log-Structured Tables in the Cloud}.
\newblock
\newblock
\showeprint[arxiv]{2305.01120}~[cs.DB]


\bibitem[\protect\citeauthoryear{Jain, Kraft, Power, Das, Stoica1, and Zaharia}{Jain et~al\mbox{.}}{2023}]%
        {lhbench}
\bibfield{author}{\bibinfo{person}{Paras Jain}, \bibinfo{person}{Peter Kraft}, \bibinfo{person}{Conor Power}, \bibinfo{person}{Tathagata Das}, \bibinfo{person}{Ion Stoica1}, {and} \bibinfo{person}{Matei Zaharia}.} \bibinfo{year}{2023}\natexlab{}.
\newblock \showarticletitle{Analyzing and Comparing Lakehouse Storage Systems}.
\newblock \bibinfo{journal}{\emph{CIDR}} (\bibinfo{year}{2023}).
\newblock


\end{thebibliography}

\end{document}